\def \SAIT #1 #2 {{\em Mem.\ Soc.\ Astron.\ It.\/} {\bf #1}, #2}
\def \MESS #1 #2 {{\em The Messenger\/} {\bf #1}, #2}
\def \ASTRNACH #1 #2 {{\em Astron. Nach.\/} {\bf #1}, #2}
\def \AAP #1 #2 {{\em Astron. Astrophys.\/} {\bf #1}, #2}
\def \AAL #1 #2 {{\em Astron. Astrophys. Lett.\/} {\bf #1}, L#2}
\def \AAR #1 #2 {{\em Astron. Astrophys. Rev.\/} {\bf #1}, #2}
\def \AAS #1 #2 {{\em Astron. Astrophys. Suppl. Ser.\/} {\bf #1}, #2}
\def \AJ #1 #2 {{\em Astron. J.\/} {\bf #1}, #2}
\def \ANNREV #1 #2 {{\em Ann. Rev. Astron. Astrophys.\/} {\bf #1}, #2}
\def \APJ #1 #2 {{\em Astrophys. J.\/} {\bf #1}, #2}
\def \APJL #1 #2 {{\em Astrophys. J. Lett.\/} {\bf #1}, L#2}
\def \APJS #1 #2 {{\em Astrophys. J. Suppl.\/} {\bf #1}, #2}
\def \APSS #1 #2 {{\em Astrophys. Space Sci.\/} {\bf #1}, #2}
\def \ASR #1 #2 {{\em Adv. Space Res.\/} {\bf #1}, #2}
\def \BAIC #1 #2 {{\em Bull. Astron. Inst. Czechosl.\/} {\bf #1}, #2}
\def \JSQRT #1 #2 {{\em J. Quant. Spectrosc. Radiat. Transfer\/} {\bf #1}, #2}
\def \MN #1 #2 {{\em Mon. Not. R. Astr. Soc.\/} {\bf #1}, #2}
\def \MEM #1 #2 {{\em Mem. R. Astr. Soc.\/} {\bf #1}, #2}
\def \PLR #1 #2 {{\em Phys. Lett. Rev.\/} {\bf #1}, #2}
\def \PASJ #1 #2 {{\em Publ. Astron. Soc. Japan\/} {\bf #1}, #2}
\def \PASP #1 #2 {{\em Publ. Astr. Soc. Pacific\/} {\bf #1}, #2}
\def \NAT #1 #2 {{\em Nature\/} {\bf #1}, #2}

\documentstyle{memsait}
\input epsf.sty
\begin{opening}
\title{THE ASCA HSS: LOOKING FOR TYPE 2 AGN} % ALL CAPITAL LETTERS PLEASE !!!
\author{Roberto Della Ceca$^1$, Valentina Braito$^2$, Volker Beckmann$^3$, 
Ilaria Cagnoni$^4$ and Tommaso Maccacaro$^1$}
\institute{
$^1$Osservatorio Astronomico di Brera, Milan, Italy\\
$^2$Osservatorio Astronomico di Padova, Padua, Italy\\
$^3$Hamburger Sternwarte, Hamburg, Germany\\
$^4$SISSA, Trieste, Italy
}
\date{} % DO NOT INSERT ANY DATE HERE !!!
\end{opening}

\begin{document}

%\oddpagefooter{\sf Mem. S.A.It., Vol. ??, ??}{}{\thepage}
%\evenpagefooter{\thepage}{}{\sf Mem. S.A.It., Vol. ??, ??}
\oddpagefooter{}{}{} % LEAVE AS IT IS !
\evenpagefooter{}{}{} % LEAVE AS IT IS !
\ 
\bigskip

\begin{abstract}
We will briefly discuss the X-ray spectral properties of the objects in
the ASCA Hard Serendipitous Survey (HSS) sample and its present AGN 
contents. We also present X-ray spectroscopy of two Type 2
AGN belonging to the sample.
\end{abstract}

\section{Introduction}

It is now largely accepted that absorbed AGN play a significant (and
perhaps major) role in the production of the Cosmic X-ray Background
(CXB) above 2 keV (see Gilli et al., these proceedings)
and should be the site where a large fraction  of the energy density of
the universe  is generated (Fabian and Iwasawa, 1999).  
However,  many important questions are still unsolved.  
For example, the relationship between narrow and broad line AGN in the
optical regime and absorbed and unabsorbed AGN in the X-ray regime is
still unclear, and doubts on the very existence of high luminosity Type 2
AGN are often cast (see e.g. Halpern, Turner and George, 1999; Akiyama
et al., 2000; Salvati and Maiolino, 2000).
We briefly address these points using the ASCA HSS sample
(Della Ceca et al., 1999a and references therein) and its present
AGN contents.  Although this small sample of identified objects is
probably not representative of the whole population, it allows us to
glimps the bright ($> 10^{-13}$ erg cm$^{-2}$ s$^{-1}, 2-10$ keV)
population of hard X-ray sources expected to be present in Chandra 
and XMM-Newton fields.
 
\section{The X-ray Spectral Properties of the ASCA HSS sample}

To investigate the X-ray spectral properties of the sources, we defined the
Hardness Ratios, $HR1 = {M-S \over M+S}$ and $HR2 = {H-M \over H+M}$,
where S, H and M are the observed net counts in the 0.7$-$2.0, 2.0$-$4.0, 
and 4.0$-$10.0 keV energy bands respectively.

In Figure 1 we report the position of the sources in the HR1-HR2 plane
for identified Type 1 and Type 2 AGN and for the complete 
ASCA HSS sample,
along with the loci expected from a single absorbed power-law model, 
with energy index and absorbing column density as indicated in the figures 
(see Della Ceca et al., 1999b for details).

\begin{figure}
\epsfysize=14cm % fix the y-dimension and scales x-dim. to y-dim.
%\epsfxsize=8cm % fix the x-dimension and scales y-dim. to x-dim.
% Feel free to do the choice you prefer but do not exceed the x-dimension
% of the text lines
\vspace{-3cm}
\hspace*{2.0cm}\epsfbox{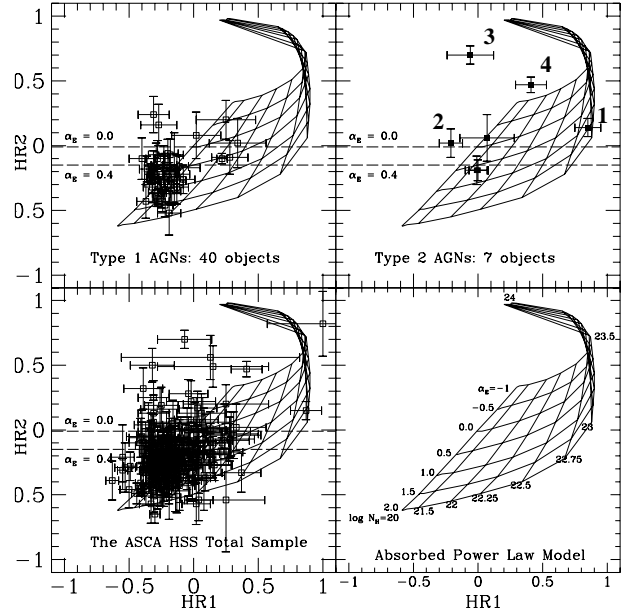} %for centering: act on hspace argument
\vspace{-3cm} 
\caption[h]{The identified Type 1 and 2 AGN and the total 
HSS sample in the HR1-HR2 plane.}
\end{figure}

It is worth noting  the large spread in HR1 and HR2 displayed by the
sources and the departure from the loci of absorbed, single power
law spectra; this implies a more complex broad band (0.7 $-$ 10 keV)
source spectrum and confirms the results previously reported by Della
Ceca et al., 1999b who used a pilot sample of 60 ASCA serendipitous
sources.  The new picture that is now emerging is that some of the optically classified
Type 1 AGN seem to have a complex and very flat ($\alpha_E \leq 0.4$)
2-10 keV spectrum, probably indicative of X-ray absorption.  A
similar result has been also obtained independently by the HELLAS team
(Vignali et al., these proceedings), based on {\it Beppo}SAX data.

%As already said in that paper a model based on the unifications scheme
%of AGNs seem to be able to explain the overall spectral properties of
%the sample.
 
\section{A Compton-Thick and a High Luminosity Type 2 AGN}

In the ``Type 2 AGN panel" of Figure 1 we have marked (n. 1 to 4) 
the 
Type 2 AGN for which we have enough counts for a more accurate 
X-ray spectroscopic analysis. 
The X-ray spectra of object \#1 and object \#2 are reported in figure
2.  Object \#1 is described by an absorbed power law model with photon
index ($\Gamma$) equal to $1.73\pm 0.5$ and absorbing column density
($N_H$) $= 5.9\pm 1.6 \times 10^{22}$ cm$^{-2}$ ($1 \sigma$
confidence errors).  This object is spectroscopically identified with a
narrow line AGN at z=0.398; at this redshift the intrinsic (i.e.
unabsorbed) 2$-$10 keV  luminosity is $\sim 2\times 10^{45}$ erg 
s$^{-1}$ 
($H_0$ = 50 km s$^{-1}$ Mpc$^{-1}$; $q_0 = 0.0$), 
making this object one of the few high luminosity Type 2 AGN
known to date.
The X-ray spectrum of object \#2 is described by the so called
``leaky-absorber" continua with $\Gamma = 1.6\pm 0.2$,
$N_H = 2.5_{-1.1}^{+1.8} \times 10^{24}$
cm$^{-2}$ and uncovered fraction of $\sim 0.8 \%$.
This object is spectroscopically identified with a nearby (z=0.046)
Seyfert 2 galaxy. Because of its high absorbing column density this object is
probably a Compton thick system with an intrinsic luminosity of
$\sim  3\times 10^{44}$ erg s$^{-1}$.
Full details on these two objects (e.g. radio, optical and X-ray   
properties) will be reported in a forthcoming paper.
Objects \#3 and \#4 are described by the ``leaky-absorber" continua
plus a narrow line(s).  
Object \#3 is  the well known Seyfert 2 galaxy NGC 6552 at z=0.026.
Its spectrum requires $\Gamma \sim 1.4$, $N_H \sim
6\times 10^{23}$ cm$^{-2}$, an uncovered fraction of $\sim 2\%$ and a
narrow Gaussian line at $\sim 6.4$ keV with an equivalent width  of 0.9
keV; its intrinsic luminosity is $\sim  6\times 10^{42}$ erg 
s$^{-1}$ (Fukazawa et al., 1994).
Finally, object \#4 was originally classified as a Type 2 AGN directly from 
X-ray data and it was subsequently confirmed as such by optical spectroscopy (Della Ceca et al., 2000). Its X-ray spectrum requires
$\Gamma = 2.5\pm 0.3$, $N_H = 1.8\pm 0.4\times 10^{23}$ cm$^{-2}$, an
unresolved Gaussian line at $\sim 6.4$ keV with equivalent width of
$\sim 0.6$ keV and uncovered fraction of $\sim 0.7\%$; at the
redshift of the object (z=0.029) its intrinsic luminosity is $\sim
1.3\times 10^{43}$ erg  s$^{-1}$.
These four examples, along with the possible existence of absorbed Type 1
AGN, show the large diversity of the sources expected to be found in the   
Chandra and XMM-Newton fields.

% To insert postscript versions of figures use the following commands:
% YOU NEED THE eps.sty FILE AND THE POSTSCRIPT FILES OF THE FIGURES
% IN THE SAME DIRECTORY.

\begin{figure}
\epsfxsize=7.0cm % fix the y-dimension and scales x-dim. to y-dim.
%\epsfxsize=8cm % fix the x-dimension and scales y-dim. to x-dim.
% Feel free to do the choice you prefer but do not exceed the x-dimension
% of the text lines
\vspace{-0.5cm}
\hspace{-0.5cm}\epsfbox{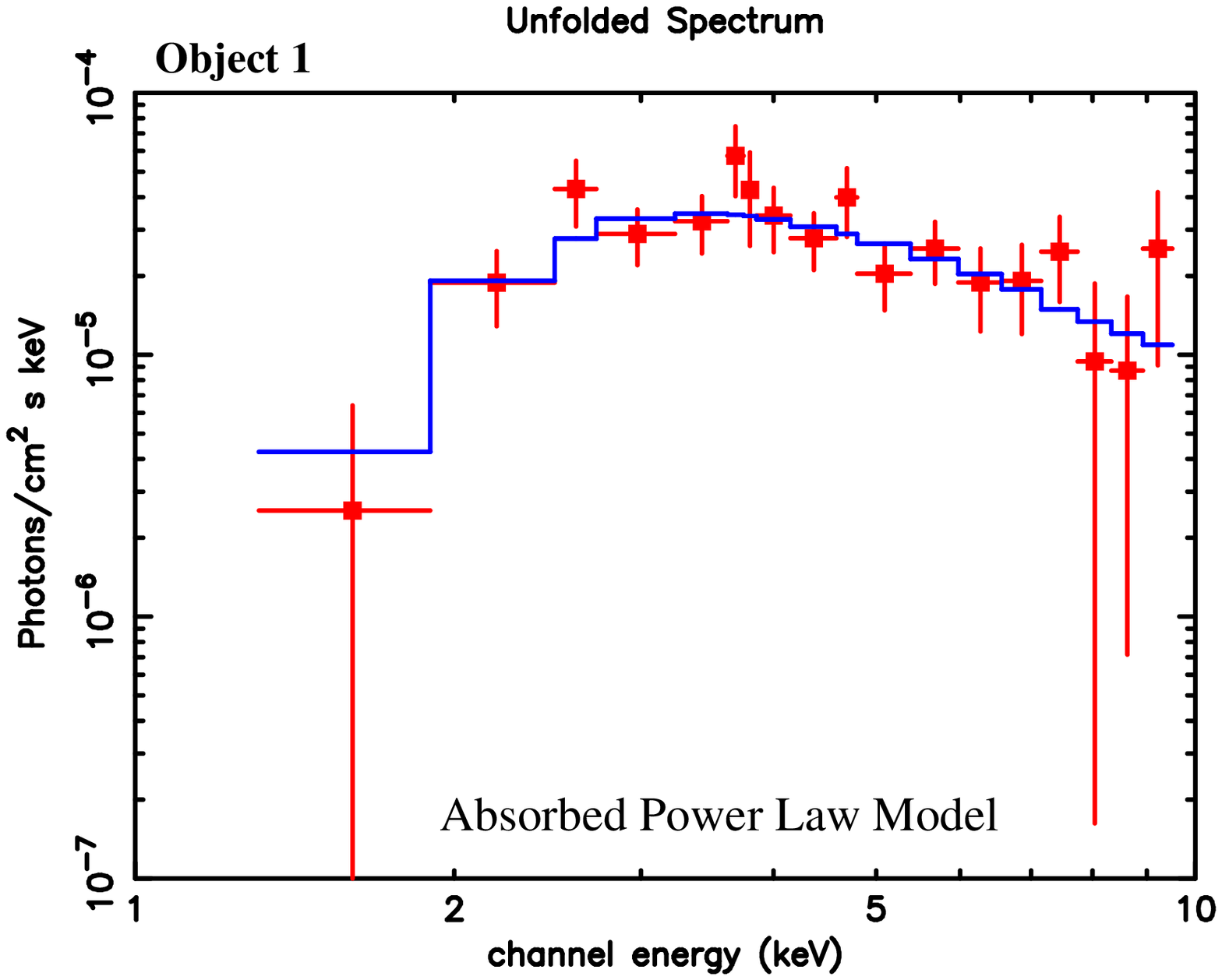}
\vspace{-5.2cm}
%\hspace{10.0cm}
\epsfxsize=7.0cm
\hspace*{7.0cm}\epsfbox{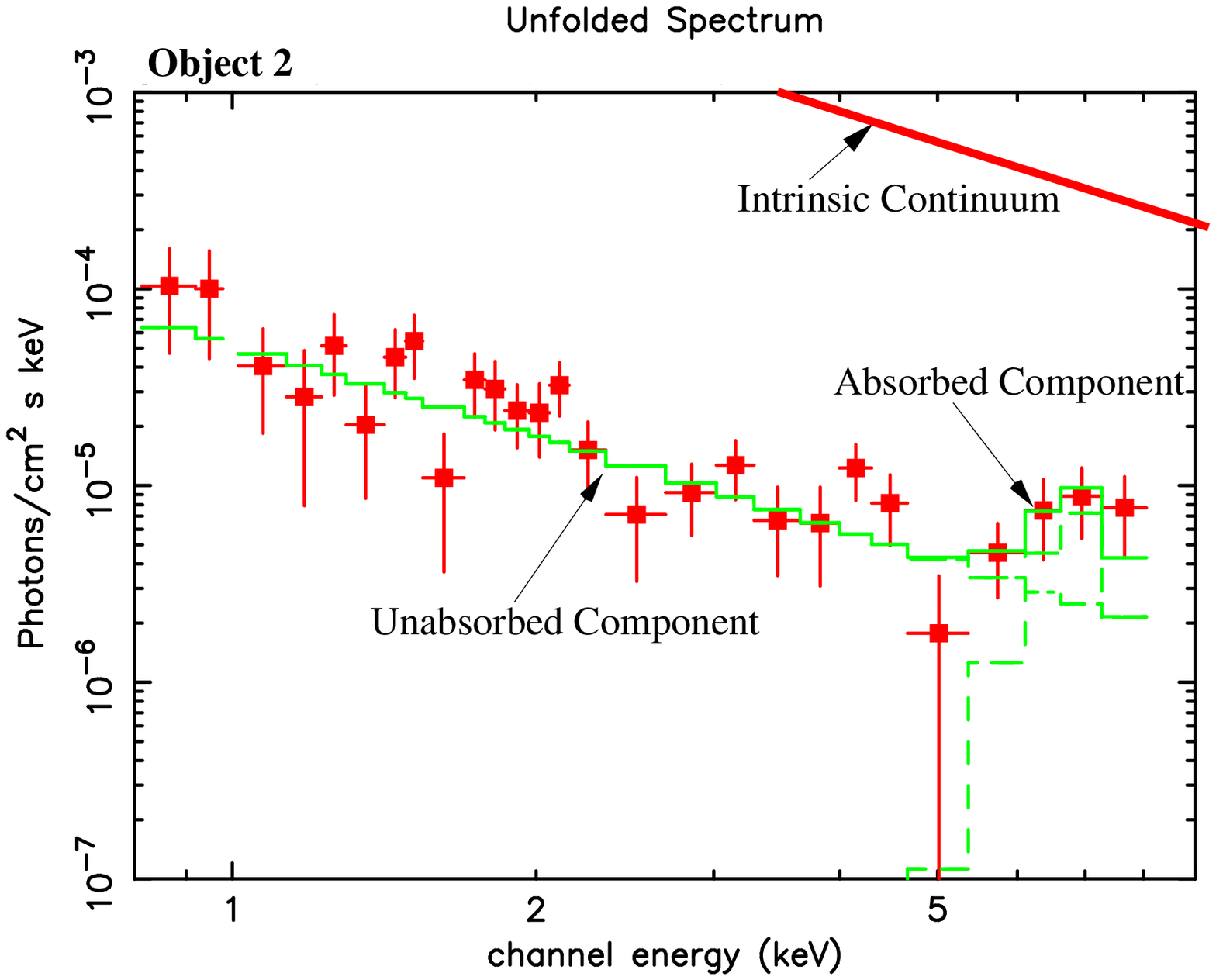} 
\caption[h]{The unfolded ASCA GIS spectra of object \#1 and object \#2.}
\end{figure}

%\vskip -2.5cm
%\begin{figure}
%\epsfxsize=7.0cm % fix the y-dimension and scales x-dim. to y-dim.
%\epsfxsize=8cm % fix the x-dimension and scales y-dim. to x-dim.
% Feel free to do the choice you prefer but do not exceed the x-dimension
% of the text lines
%\vspace{-0.1cm}
%\hspace{-1.5cm}\epsfbox{a1511_xspec.ps}
%\vspace{-5.0cm}
%\hspace{10.0cm}
%\epsfxsize=7.0cm
%\hspace*{7.0cm}\epsfbox{obj9593g100.ps2} 
%\caption[h]{Type here the caption for the figure.}
%\end{figure}

% To insert harcopy versions of figures use the following commands:
% FIGURES MUST BE MOUNTED IN THE APPROPRIATE SPACES BEFORE
% SUBMITTING THE CAMERA-READY HARDCOPY.
%\begin{figure}
%\vspace{2cm}   % empty space for the figure (2cm in the given example)
%\caption[h]{Type here the caption for the figure.}
%\end{figure}

\acknowledgements
This work received partial financial support from MURST under grant Cofin98-02-32.

% References. We avoided using the \bibitem commmand since we found it is
% somewhat platform-dependent. We also avoided using the \cite{keyword}
% command since we found it cumbersome. However, if you are an expert 
% LateX user you may use the various LateX tools for the references 
% provided they give the same printout formats of the examples given here.

\end{document}